\documentclass[]{spie}  

\newcommand{\OVI}{\hbox{{\rm O}\kern 0.1em{\sc vi}}}
\usepackage{amsmath,amsfonts,amssymb}
\usepackage{graphicx}
\usepackage[colorlinks=true, allcolors=blue]{hyperref}
\usepackage{tabularray}
\usepackage{tabularx}
\usepackage{makecell}
\usepackage{setspace}
\usepackage{multirow}
\usepackage{multicol}
\usepackage{geometry}
\usepackage{tabularray}
\UseTblrLibrary{booktabs}
\usepackage[T1]{fontenc}
\usepackage[utf8x]{inputenc} 
\usepackage{subfigure}
\usepackage[subfigure]{tocloft}
\usepackage{subcaption}
\usepackage{graphicx} 
\usepackage{tablefootnote}

\newcommand\electron{e\textsuperscript{-}}
\cftpagenumbersoff{figure}
\cftpagenumbersoff{table}

\title{Advancing Ultraviolet Detector Technology for future missions: Investigating the dark current plateau in silicon detectors using photon-counting EMCCDs}

\author[a,d]{Aafaque R. Khan}
\author[a]{Erika Hamden}
\author[b]{Gillian Kyne}
\author[b]{April D. Jewell}
\author[b]{John Henessey}
\author[b]{Shouleh Nikzad}
\author[c]{Vincent Picouet}
\author[a]{Olivia Jones}
\author[a]{Harrison Bradley}
\author[a]{Nazende Kerkeser}
\author[c]{Zeren Lin}
\author[a]{Brock Parker}
\author[a]{Grant West}
\author[a]{John Ford}
\author[a]{Frank Gacon}
\author[a]{Dave Beaty}
\author[a]{Jacob Vider}
\affil[a]{Steward Observatory, University of Arizona, 933 N Cherry Avenue, Tucson, Arizona, United States of America}
\affil[b]{Jet Propulsion Laboratory, California Institute of Technology, 4800 Oak Grove Dr, Pasadena, California, United States of America}
\affil[c]{California Institute of Technology, 1200 E California Blvd, Pasadena, California, United States of America}
\affil[d]{Wyant College of Optical Sciences, University of Arizona, 1630 E University Blvd, Tucson, Arizona, United States of America}
\authorinfo{Further author information: (Send correspondence to Aafaque R. Khan)\\A.R.K.: E-mail: arkhan@arizona.edu, Telephone: 1 520 223 5171}
\begin{document} 
\maketitle

\begin{abstract}
Understanding the noise characteristics of high quantum efficiency silicon-based ultraviolet detectors, developed by the Microdevices Lab at the Jet Propulsion Laboratory, is critical for current and proposed UV missions using these devices.  In this paper, we provide an overview of our detector noise characterization test bench that uses delta-doped, photon counting, Electron-multiplying CCDs (EMCCDs) to understand the fundamental noise properties relevant to all silicon CCDs and CMOS arrays. This work attempts to identify the source of the dark current plateau that has been previously measured with photon-counting EMCCDs and is known to be prevalent in other silicon-based arrays. It is suspected that the plateau could be due to a combination of detectable photons in the tail of blackbody radiation of the ambient instrument, low-level light leaks, and a non-temperature-dependent component that varies with substrate voltage. Our innovative test setup delineates the effect of the ambient environment during dark measurements by independently controlling the temperature of the detector and surrounding environment. We present the design of the test setup and preliminary results.

\end{abstract}

\keywords{delta-doped EMCCDs, dark current plateau, Ultraviolet spectrograph, detector noise performance,delta-doped CCD, detector optimization}

    

\section{INTRODUCTION}
\label{sec:intro} 
aSilicon-based CCD and CMOS detectors processed with delta-doping\cite{Hoenk_1992,Nikzad_1994,Nikzad_2012}, a Molecular Beam Epitaxy (MBE) process developed at the Microdevices Laboratory (MDL) at the Jet Propulsion Lab (JPL), have been demonstrated to achieve Internal Quantum Efficiency (IQE) near 100\% in the Optical/Ultraviolet (UV) spectrum.\cite{Nikzad_2017,Kyne16, Hamden_2012} Combined with Atomic Layer Deposition (ALD) anti-reflection (AR) coatings these devices have greater than 60\% Quantum Efficiency (QE) in the UV\cite{Hennessy_2017,Hennessy:15, Hamden_2012}. Several ongoing and upcoming missions (e.g. FIREBall-2 \cite{Hamden_2020}, SPARCS \cite{Gamaunt_2022}, UVEX \cite{kulkarni2023science}) and instrument concepts (Hyperion \cite{hamden2022hyperion}, UV-Scope \cite{UVScope2022}, Eos \cite{hamden:2023}, Nox \cite{Chung:2022}, and others) are leveraging these detectors, combined with advanced UV mirror coatings, to open new frontiers in UV Astronomy. A key focus in the development of silicon-based detector technology for future missions, including ultraviolet instrument concepts for the Habitable Worlds Observatory, is the characterization of their noise performance under different operating conditions and surrounding environments.

Electron-Multiplying Coupled Devices (EMCCDs) \cite{LLCCD_Jerram,EMCCD_Craig} have emerged as an exciting detector option in astrophysics due to their photon-counting capability and low read noise ($\ll$ 1e-/pixel/frame in photon counting mode){\cite{Tulloch_2011,Kyne16}}. These detectors are specifically favorable for photon-starved UV astronomy where the low background (an order of magnitude lower than visible) allows high-sensitivity measurements that are typically limited by the detector noise and instrument background \cite{Martin2019IGM}. The first sub-orbital demonstration of a UV-enhanced, delta-doped, ALD-coated EMCCD detector was flown on the Faint Intergalactic Red-shifted Emission Balloon (FIREBall-2, hereafter FB2) in 2018 \cite{Hamden_2020}. Similar non-delta-doped devices that are optimized for Optical/Near-Infrared will fly on the Coronagraphic Instrument (CGI) of the Nancy Grace Roman Space Telescope \cite{Morrissey_2023}.  

In this paper, we provide an overview of our detector noise characterization test bench that uses delta-doped, photon counting, EMCCDs to understand the fundamental noise properties relevant in all silicon CCDs and CMOS arrays. The focus of this work is to identify the source of the dark current plateau that limits the performance of EMCCDs in photon-counting mode (Sec. \ref{sec:dark_curent_plateau}). While it is expected that the thermal dark current in silicon detectors should decrease exponentially with temperature \cite{Janesick_1990,Widenhorn_2022}, previous dark measurements with EMCCDs have shown to plateau below a temperature of 160 K (Sec.\ref{sec:dark_curent_plateau}). We report the design of a novel test setup (Sec. \ref{sec:Experimental setup}) to further investigate the dark plateau using delta-doped EMCCDs operated with noise-optimized front-end and readout electronics (Sec. \ref{subsect:detector_controller}). The setup is designed to measure the dark current in a controlled environment with low-light leaks. The detector temperature is precision controlled while the temperature of the surrounding environment of the detector is controlled independently (Sec \ref{subsec:Test setup design}). To take dark exposures over several weeks, we have automated the test bench for the uninterrupted acquisition of dark exposures in different configurations of the test setup (Sec \ref{subsec:concept_of_operation}). We present our preliminary results from the first dark current measurements with different combinations of temperatures for the detector and its ambient environment in section \ref{sec:measurements}. The preliminary measurements presented here are part of our ongoing effort to understand the dark current in silicon-based detectors (Sec. \ref{sec:current_future}) and develop operational and system design strategies to improve their performance.


\section{THE DARK CURRENT PLATEAU}
\label{sec:dark_curent_plateau}
The single photon counting capability of EMCCDs makes them an excellent tool for the precision measurement of noise characteristics of similar silicon CCD and CMOS arrays. The EMCCD architecture overcomes the low-frequency readout noise, the limiting noise in conventional CCDs, by amplifying the signal in the serial gain registers using avalanche multiplication \cite{Hynecek_2001}. EMCCDs can accommodate a wide dynamic range of signals by operating with a variable gain (from $>$1,000  e$^{-}$/e$^{-}$ in photon-counting mode to 1  for conventional CCD mode \cite{Olivier_2008}). But with this pre-amplifier gain approach, the noise due to dark current and Clock Induced Charge (CIC), which is normally negligible compared to read noise in traditional CCDs, is also amplified like the photo-electrons. Furthermore, the avalanche multiplication process generates additional noise from cosmic ray trails, deferred charge, and coincidences \cite{Kyne_2020,Olivier_2008}.

Previous efforts to optimize the noise performance of these devices including improving the readout electronics \cite{Daigle_2009,Daigle_2012,Daigle_2022}, detector cooling, innovative clock shaping \cite{Daigle_2012,Kyne16,Bush_2021}, and modifications to the device architecture \cite{Morrissey_2023} have resulted in a significant reduction in the noise floor. The CIC and dark current remain the noise-limiting factors in the performance of these devices \cite{Picouet_2022}. The dark current in silicon devices is expected to decrease exponentially with detector temperature \cite{janesick2001scientific,Widenhorn_2022}. However, previous measurements with EMCCDs have shown it to plateau around a detector temperature of 163 K \cite{Kyne2020,Kyne16}. It is suspected that this plateau could be due to a combination of ambient thermal blackbody VIS/IR photons from the dewar, low-level light leaks, and/or a non-temperature dependent component, that varies with substrate voltage, caused by spurious charge generation in the detector pixels and/or emission from the detector packaging. Similar plateaus have been observed in other CCDs and an investigation to understand this noise will have a wide and profound impact on understanding low-noise CCDs and CMOS arrays of all types.

We hypothesize that the dark plateau measured in the EMCCDs is not caused by the intrinsic thermal dark current in the pixels and is from the visible photons in the tail of thermal black-body emission from the surrounding hardware. Unlike infrared telescopes and instruments, where both the detectors and the telescope structures are cooled, UV and optical instruments do not employ cooling of the mechanical surrounding structure of the telescope and camera. To test this hypothesis, we employ two independent cooling chains to cool the detector and the surrounding environment and measure the dark current at different combinations of temperatures and substrate voltages.

\begin{figure}[htb]
    \centering
    \includegraphics[width=1\textwidth]{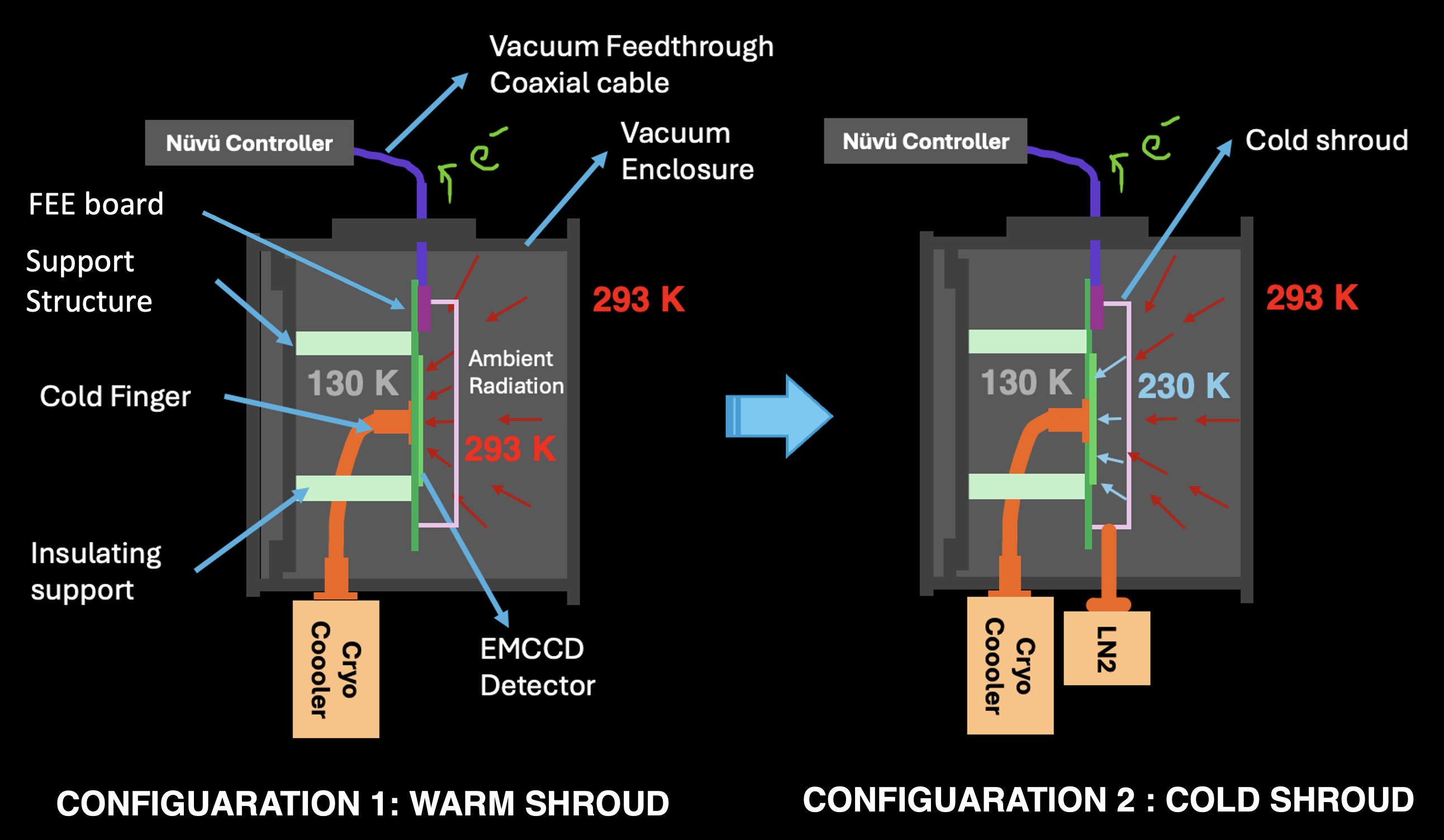}
    \caption{Dark Current measurement Configurations. (Left) In the warm shroud configuration, the surrounding shroud is at room temperature while the detector is cooled to different temperatures for the dark current measurements while in the Cold Shroud configuration (Right) the temperature of the shroud is controlled at around 230 K with Liquid Nitrogen (LN2) shroud and patch heaters.}
    \label{fig:configurations}
\end{figure}

\section{Experimental setup}
\label{sec:Experimental setup}
\subsection{Test setup design}
\label{subsec:Test setup design}
Our novel detector characterization test setup allows us to control the detector and the surrounding temperature independently. The thermal control system has been designed with control authority to operate the detector at any temperature between 140 K to 180 K, while the shroud can be kept at any temperature between room temperature and 170 K. In this section, we provide a summary of the test setup design. The gain of the serial multiplication register is very sensitive to the temperature of the detector. Therefore, the detector cooling chain has been designed to achieve precision control of detector temperature to within $\pm0.5$ K. A detailed description of the design of the test setup will be provided in Khan et al. (in preparation)\cite{Khan_2024}.

\begin{figure}[htb]
    \centering
    \includegraphics[width=1\textwidth]{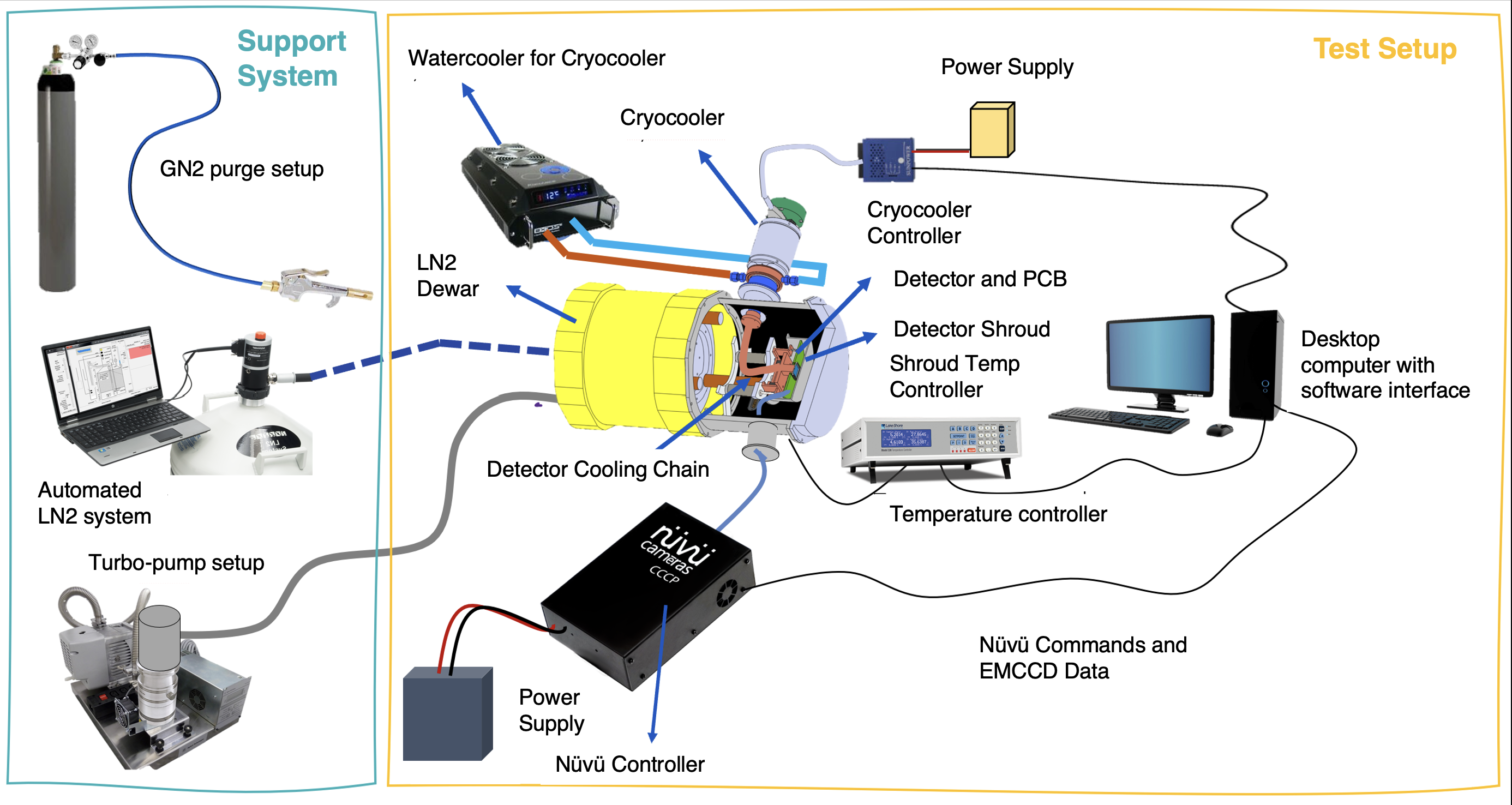}
    \caption{Schematic of the test setup, illustrating the vacuum dewar, detector and detector controller, detector and shroud cooling systems, and ancillary components including the water cooling for the cryocooler, vacuum pump system, automatic LN2 refill system, and ultra-high purity gaseous Nitrogen supply for purging.}
    \label{fig:setup_schematic}
\end{figure}

The schematic of the dark characterization test setup is shown in Figure \ref{fig:setup_schematic}. The detector and front-end electronics are housed in a high vacuum enclosure consisting of an ND8 dewar from IRlabs and a custom-built extension. The vacuum enclosure is evacuated by a TPS-flexi vacuum pumping system from Agilent\footnote{Agilent Technologies, Inc., 5301 Stevens Creek Blvd, Santa Clara, CA 95051, USA}. A full-range Pirani/Inverted Magnetron pressure gauge is used to monitor the pressure of the system from the turbo controller. The detector is connected to the cold tip of a Cryotel\footnote{Sunpower Inc.m 2005 E. State St. Suite 104, Athens, Ohio 45701, USA} MT Crycoocoler with 15-20 W of lift between 140-180 K (at 298 K reject temperature). The cryocooler is controlled by a closed-loop controller that monitors a Pt-100 temperature sensor on the cold tip to maintain it at a fixed programmable temperature. A Lakeshore\footnote{Lakeshore Cryotonics, 550 Tressler Dr, Westerville, OH, USA} Model 350 controller provides the precision temperature control of the detector with patch heaters and temperature sensors mounted in proximity to the detector. The detector is surrounded by a shroud that is connected to the LN2 cold plate of the dewar. Any direct line of sight from the image area points to the surface of the shroud. For cooling the shroud, the dewar is filled with LN2 and the Lakeshore controller uses a Pt-100 sensor and patch heaters on the shroud to control its temperature. The LN2 level in the dewar is maintained by a Norhof\footnote{Norhof LN2 microdosing systems, Galileïlaan 33U 6716BP Ede, The Netherlands} automated LN2 micro-dosing system to allow multi-day operations without a manual LN2 refill. The control for the detector and shroud cooling chains is automated with a Python\cite{python} script that interfaces with the cryocooler controller and Lakeshore controller to monitor and control the temperatures.

The dark measurements for this work are taken in two different configurations, as shown in figure \ref{fig:configurations}, with the shroud at ambient temperature (293 K) and the shroud cooled to cryogenic temperature (230 K).


\subsection{Detector and controller}
\label{subsect:detector_controller}
\begin{figure}[htb]
    \centering
    \includegraphics[width=0.8\textwidth]{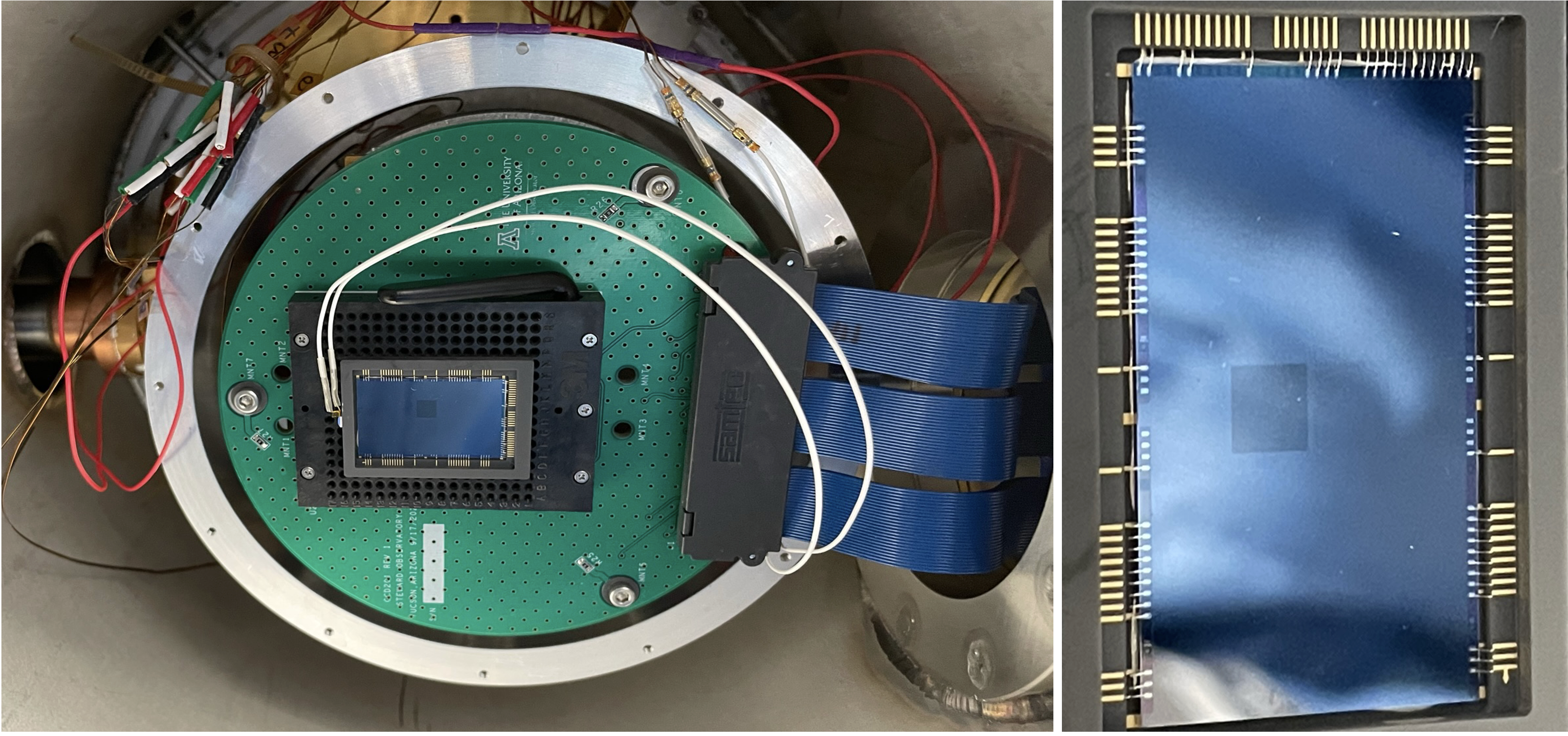}
    \caption{EMCCD used for dark measurements. (Left) The EMCCD is mounted on the front-end electronics board inside the dewar. The Samtech cable for vacuum feedthrough is on the right of the image. The detector shroud was removed to access the detector. (Right) Close-up of the custom delta-doped EMCCD used for dark measurements. This custom Te2v 201-20 EMCCD was delta-doped and ALD coated at the JPL Microdevices Laboratory (See Jewell et al.).\cite{April_2023}}
    \label{fig:Detector}
\end{figure}

The detector used for the measurements reported in this work is a CCD201-20 EMCCD (Teledyne-e2v) that has been through backside processings, including delta doping at the MDL\cite{April_2023} (Figure \ref{fig:Detector}). It is mounted on a custom front-end electronics board that interfaces with the N\"uv\"u CCCP v3 controller\cite{Daigle_CCCP_2008}. This board has been configured for the 1 MHz high-voltage clock readout. The front-end electronics board for the design was adopted from the FB2 flight boards \cite{Lingner2017,Hamden_2020}. Significant changes were made to the original design by optimizing the track layout, separating the bias and clock tracks into different layers, and adding grounding planes on the board. This optimization improved the noise performance of the detector signal chain achieving $\sim$ 18 \electron{} rms read noise (see section \ref{subsec:concept_of_operation}). Additionally, the SIP (Single In-line Package) receiving sockets mounting the detector were replaced with ZIF (Zero Insertion Force) sockets to enable a more reliable and faster replacement of detectors in the setup. 

\begin{figure}[htb]
    \centering
    \includegraphics[width=0.85\textwidth]{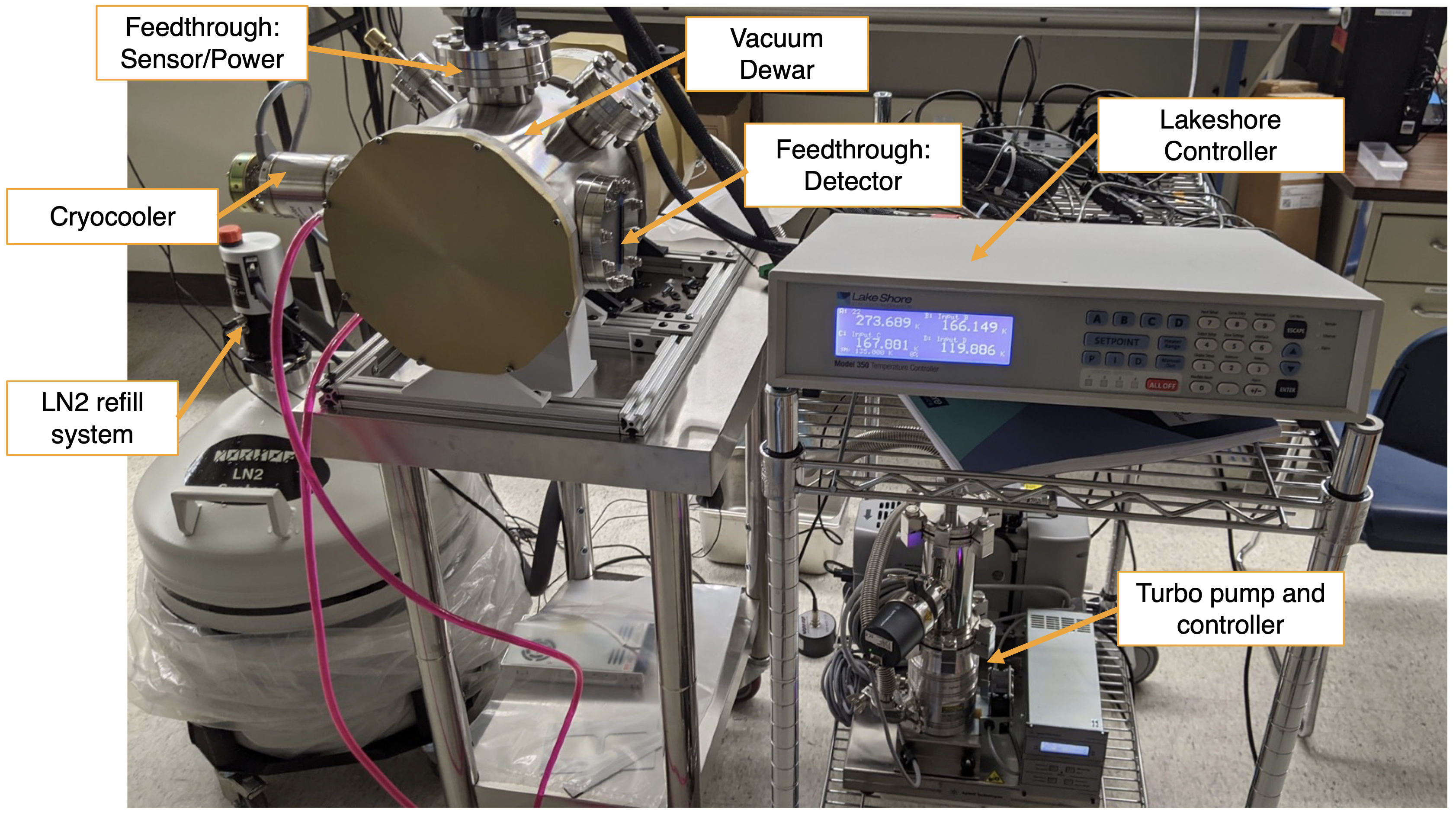}
    \caption{Schematic of the test setup, illustrating the vacuum dewar, detector and detector controller, detector and shroud cooling systems, and ancillary components including the water cooling for the cryocooler, vacuum pump system, automatic LN2 refill system, and Ultra-High Purity Gaseous Nitrogen supply for purging.}
    \label{fig:plot}
\end{figure}

\subsection{Concept of Operation}
\label{subsec:concept_of_operation}

Before making dark measurements, the conversation gain and read noise are measured by taking Photon Transfer Curves (PTC) in the conventional CCD mode with multiplication gain (G) set to 1 \electron{}/\electron{}\cite{janesick_photon_transfer}. 
The details of our PTC measurement setup will be discussed in Khan et al (in prep) \cite{Khan_2024}. Briefly, the detector is illuminated with a uniform monochromatic beam from a custom vacuum monochromator, and several exposures with progressively increasing exposure duration are taken with the detector. These exposures are used to create a PTC to extract the conversion gain and read noise. For the detector used in this test setup, 8 independent PTC measurements were taken at VSS= 0 V. The average read noise ($\sigma$) was found to be $\sim$ 18 \electron{} rms with an average conversion gain of 1.382 \electron{}/ADU. For each VSS setting at which dark current is measured, the read noise and conversion gain must be measured using the PTC.

The dark current measurements are taken in a controlled environment to minimize light leaks. The setup is in a laboratory room prepared to block almost any ambient light sources. Additionally, the dewar is made light-tight with no viewing windows or ports. The end-to-end acquisition is executed through a Python app that can be configured to run exposure sequences with different values of controller parameters, detector, and shroud temperature settings. 

Dark measurements taken for this work are summarized in Table \ref{tab:dark_data}. The dark exposures were taken for 10 different combinations of the detector (180 K to 130 K in steps of 10 K) and shroud temperatures (298 K and 230 K). For each of these combinations, five dark and five bias exposures were taken for 15 different exposure settings between 0 to 600 seconds. We took 3750 images with a total exposure and readout duration of $\sim$30 hours. All these images are for one substrate voltage setting (VSS=0) and we are in the process of collecting similar datasets for different VSS settings. 

The detector was operated in photon counting mode, with the multiplication gain kept at $\sim$1000 \electron{}/\electron{}. The amplitude for the High Voltage (HV) clock was adjusted in the readout sequence to keep the fixed multiplication gain for different detector operational temperatures. Due to the stochastic nature of multiplication, the calibration of the HV clock voltages for each temperature was done by taking several short-duration measurements while adjusting the HV DAC value till the mean EMgain was $\sim$1000 \electron{}/\electron{}. A high multiplication gain to achieve 
G/$\sigma$ > 50 ensures that >90 $\%$ of events are counted when a 5.5$\sigma$ cut-off is used for PC threshold \cite{Daigle_2010}.

\begin{table}
\centering
\caption{Summary of the Temperature and Acquisition Parameters for Dark Measurements. The dark rate was measured at different detector temperatures with shroud at 298 K and 230 K. Dark images were acquired with exposure ranging from 0 to 600 seconds. For each detector temperature setting the HV DAC value was set to get a constant EMgain of 1000 \electron{}/\electron{}. The substrate voltage (VSS) was set to 0V. Measurements for different VSS settings between 0-6V are ongoing.  }
\label{tab:dark_data}
\begin{tblr}{
    colspec=
    {
    Q[c,wd=0.04\linewidth]*
    {2}{Q[c,wd=0.12\linewidth]}
    Q[c,wd=0.045\linewidth]
    Q[c,wd=0.12\linewidth]
    Q[c,wd=0.12\linewidth]
    Q[c,wd=0.1\linewidth]
    Q[c,wd=0.14\linewidth]
    },
    row{1}={font=\bfseries},
}
\toprule
\SetCell[r=2]{c}{Test ID} & \SetCell[r=2]{c}{Detector Temp. [K]} & \SetCell[r=2]{c}{Shroud Temp. [K]} & \SetCell[r=2]{c}{VSS [V]} & \SetCell[r=2]{c}{Exposure Time Range [s]} & \SetCell[r=2]{c}{HV DAC value} & \SetCell[r=2]{c}{Darks per exp. setting} \\
& & & & & & \\
\midrule
\SetCell[c=7]{c}{\textbf{Warm Shroud Configuration}}\\
\midrule
1 & 183 $\pm$ 0.3 & 298 & 0 & 0-600 & 8176 & 5 \\
2 & 173 $\pm$ 0.3& 298 & 0 & 0-600 & 7990 & 5 \\
3 & 163 $\pm$ 0.3& 298 & 0 & 0-600 & 7820 & 5 \\
4 & 153 $\pm$ 0.3& 298 & 0 & 0-600 & 7795 & 5 \\
5 & 143 $\pm$ 0.3& 298& 0 & 0-600 & 7780 & 5 \\
\midrule
\SetCell[c=7]{c}{\textbf{Cold Shroud Configuration}}\\
\midrule
6 & 183 $\pm$ 0.3& 230 & 0 & 0-600 & 8176 & 5 \\
7 & 173 $\pm$ 0.3& 230 & 0 & 0-600 & 7990 & 5 \\
8 & 163 $\pm$ 0.3& 230 & 0 & 0-600 & 7820 & 5 \\
9 & 153 $\pm$ 0.3& 230 & 0 & 0-600 & 7795 & 5 \\
10 & 143 $\pm$ 0.3& 230 & 0 & 0-600 & 7780 & 5 \\
\bottomrule
\end{tblr}
\end{table}

\section{PRELIMINARY DARK CURRENT MEASUREMENTS AND ANALYSIS}
\label{sec:measurements}
\begin{figure}[htb]
    \centering
    \includegraphics[width=1\textwidth]{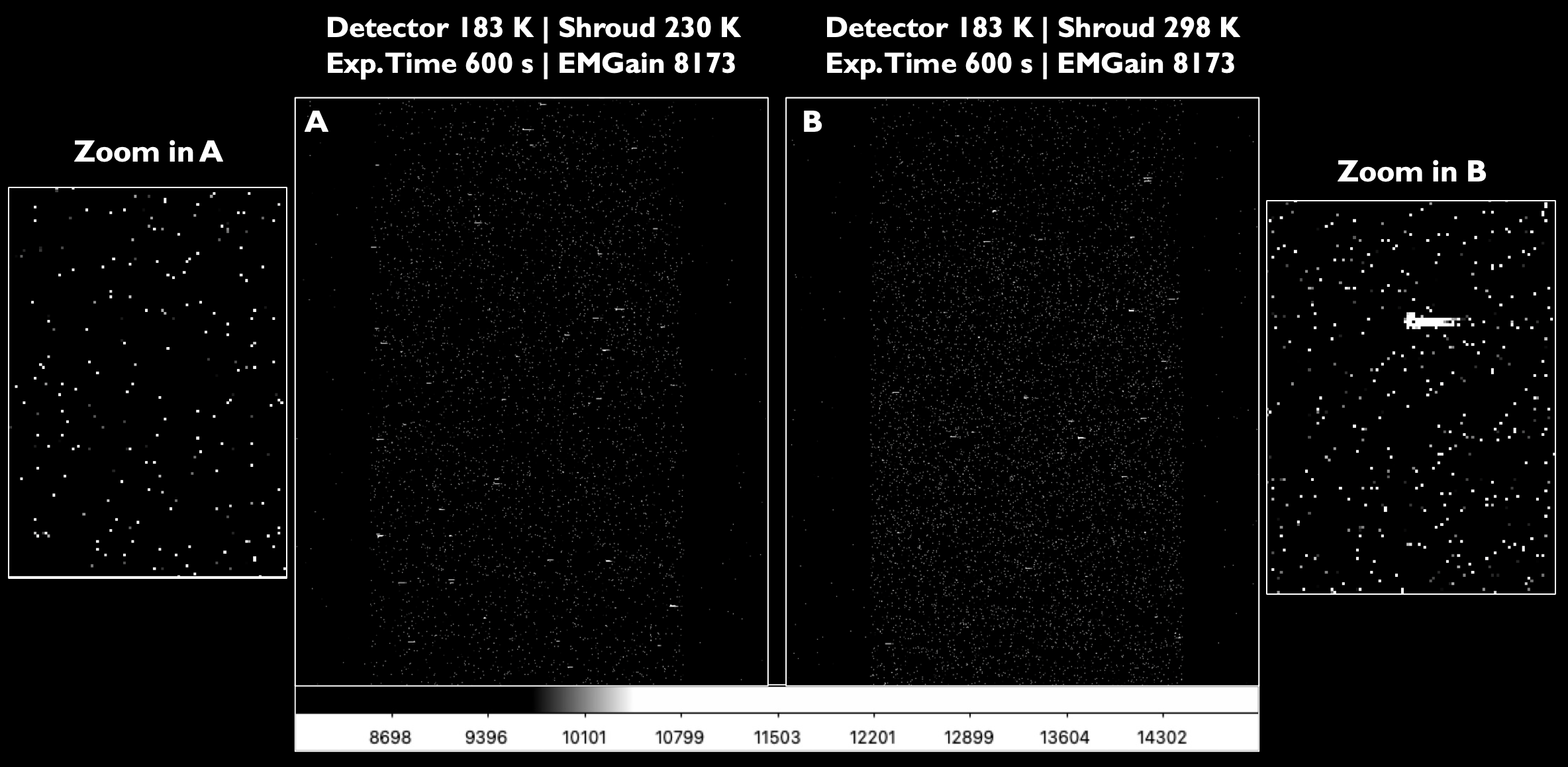}
    \caption{Raw Dark images with 600 second exposure time at 183 K in the warm shroud (A) and cold shroud (B) configurations. The 1024x2048 pixels of the image and storage area of the detector can be identified distinctly from the pre-scan and over-scan regions. The prescan and overscan regions are the same size as the detector pixels but only a portion of those regions is shown. The zoom-in view shows the distinct pixels. At 183 K there is no noticeable smearing. The zoom-in view for Image B has a cosmic ray event with a distinct tail formed due to the over-spilling of electrons generated by cosmic ray strike in the multiplication register.}
    \label{fig:dark_example}
\end{figure}

In this section, we present our preliminary dark current measurements with the warm and cold shroud configurations. We took several datasets with the setup to streamline the operations of cooling chains and detector acquisition, optimizing readout sequences, and mitigating light leaks. The results presented here are from the first complete set of measurements at a substrate voltage of 0 V taken with this test bench. Figure \ref{fig:dark_example} shows an example of raw dark images taken at 183 K in the two shroud temperature configurations.  

\subsection{Data reduction in PC Mode}
\label{subsec:DarkPC_reduction}
The raw dark images were reduced using a Python-based photon counting dark reduction pipeline described in Kyne et al\cite{Kyne2020}. The reduction process is summarized here. The data for the detector and shroud temperature combination is reorganized by collecting all images with the same exposure time into an image cube. For each image cube,  the hot pixels, cosmic ray events, and cosmic ray tails are masked and replaced. To replace the marked pixels, a random Gaussian noise array is created that has the same size as the image. The standard deviation of the Gaussian is the same as the read noise in each image. It is estimated by fitting a Gaussian around the bias peak in the histogram of each image. The replacement value for masked pixels is picked from the corresponding pixels in the Gaussian noise array. The corrected image cubes are saved as separate FITS files. 

The next step is desmearing the images. At low temperatures, a reduction in Charge Transfer Efficiency (CTE) leads to the smearing of photoelectrons generated in one pixel to the neighboring pixels as the charge is shifted during readout \cite{Daigle_CCCP_2008}. A histogram of each corrected image is generated to distinguish read noise from potential events using appropriate thresholds. Values of the trailing pixels of the potential event pixel are added to the event pixel until the trailing pixels are above a selectable smear threshold. The trailing pixels above the threshold are replaced with values from the Gaussian noise array. Finally, a histogram of each cosmic ray, hot-pixel, and the smear-corrected image is used to find the value for the 5.5$\sigma$ threshold\cite{Daigle_2010}. A binary image is generated with all pixels above the threshold value identified as events and assigned a value of 1 while all other pixels are set to 0. The events in the thresholded images are a combination of dark, parallel, and serial CIC events.

\begin{figure}[ht]
    \centering
    \includegraphics[width=1\textwidth]{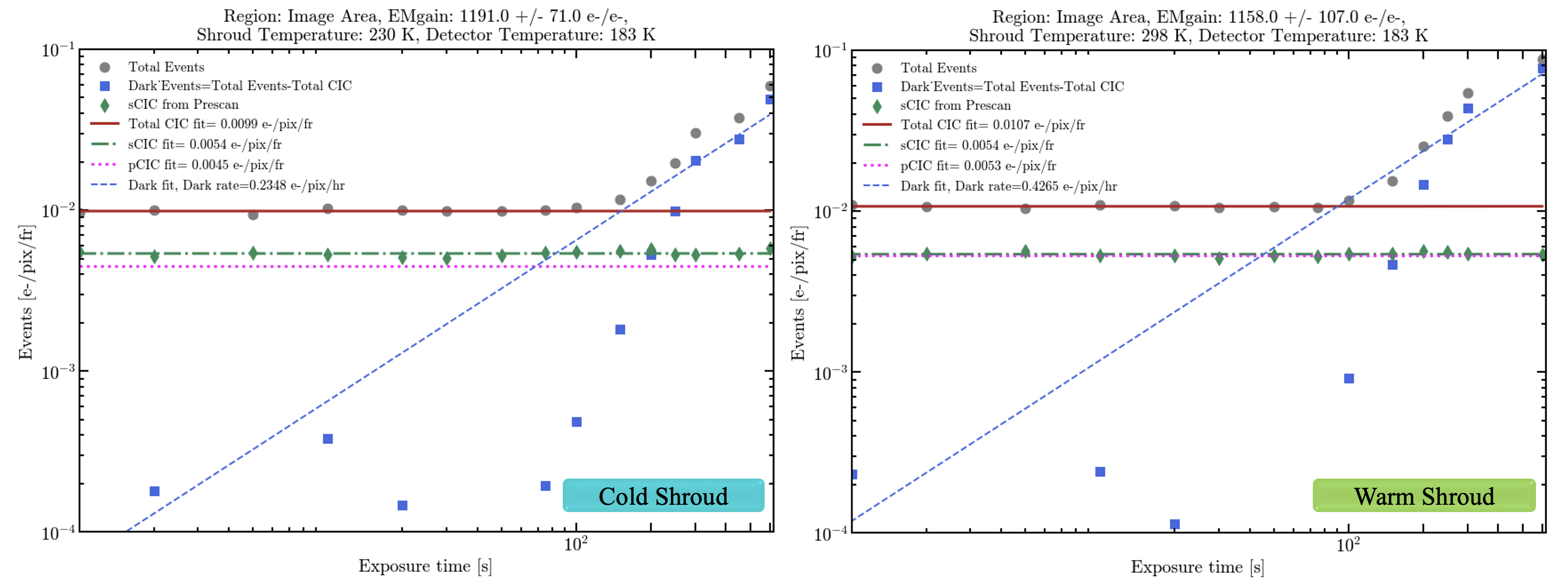}
    \includegraphics[width=1\textwidth]{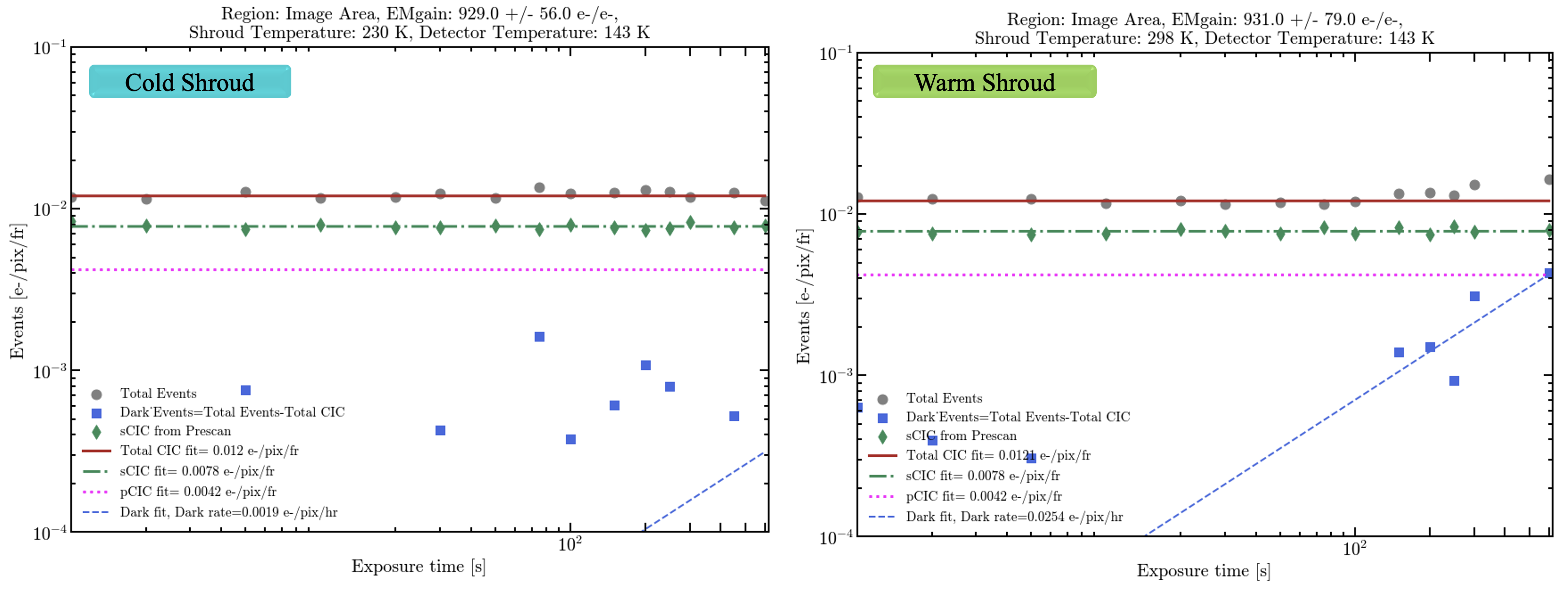}
    \caption{Photon transfer curves for the dataset with detector temperature at 183 K (Top) and 143 K (Bottom) for the two shroud temperatures.}
    \label{fig:dark_ptc_1}
\end{figure}

\subsection{Dark Photon Transfer Curves}
\label{subsec:dark_ptc}
For each detector temperature, a Dark Photon Transfer Curve (Dark-PTC) is created from the thresholded dark images by plotting the total events vs exposure duration. While the dark rate increases with exposure time, the number of total CIC events is not dependent on exposure time. The mean number of total events from each image cube (five images at each exposure time value) is plotted against the exposure time to generate a Dark-PTC. The total counts in low-exposure images are dominated by total CIC while the contribution from the dark events increases with increasing exposure. The serial CIC is estimated from the prescan region. The parallel CIC is estimated by subtracting the serial CIC (sCIC) from the total CIC. The dark rate is the slope of a line fit to the difference of total events from total CIC. Figure \ref{fig:dark_ptc_1} shows examples of Dark-PTCs for the images with the detector at 183 K and 143 K in the two shroud configurations.


\subsection{Detector Performance}

Table \ref{tab:dark_results} provides a summary of the measured EMGain, serial, parallel, \& total CIC, and dark rate for each detector and shroud temperature configuration (Table \ref{tab:dark_data}). The following sections provide a detailed description and analysis of the results. 

\begin{table}[ht]
    \centering
    \caption{Sample of the multiplication gain and noise measured in the "Image area" for the tests listed in Table \ref{tab:dark_data}. We report the average value of EM gain with 1$\sigma$ deviation, CIC noise is measured from the bias and prescan data, and the dark rate is estimated by fitting a curve to the image area noise vs exposure time.}
    \label{tab:dark_results}
    \begin{tblr}{
        colspec={
        Q[c,m,wd=0.05\linewidth]
        Q[c,m,wd=0.16\linewidth]
        Q[c,m,wd=0.15\linewidth]
        Q[c,m,wd=0.15\linewidth]
        Q[c,m,wd=0.15\linewidth]
        Q[c,m,wd=0.15\linewidth]},
        row{1}={font=\bfseries},
    }
    \hline
    \SetCell[r=2]{c}{Test ID} & \SetCell[r=2]{c}{EM gain [\electron{}/\electron{}] } & \SetCell[r=2]{c}{Serial CIC [\electron{}/pix/frame]} & \SetCell[r=2]{c}{Parallel CIC [\electron{}/pix/frame]} & \SetCell[r=2]{c}{Total CIC [\electron{}/pix/frame]} & \SetCell[r=2]{c}{Dark Rate [\electron{}/pix/hr]} \\
    & & & & & \\
    \midrule
    \SetCell[c=6]{c}{\textbf{Warm Shroud Configuration}}
    \\
    \midrule
    1 & 1095.6$\pm$178.7 & 5.5 x $10^{-3}$ & 5.3 x $10^{-3}$ & 1.07 x $10^{-2}$& 4.53 x $10^{-1}$ \\
    2 & 1147.3$\pm$103.8 & 4.0 x $10^{-3}$ & 5.2 x $10^{-3}$ & 9.0 x $10^{-3}$& 6.74 x $10^{-2}$ \\
    3 & 1098.2$\pm$131.4 & 3.7 x $10^{-3}$ & 4.7 x $10^{-3}$ & 8.5 x $10^{-3}$& 1.36 x $10^{-2}$ \\
    4 & 1059.9$\pm$160.2 & 4.3 x $10^{-3}$ & 5.9 x $10^{-3}$ & 1.02 x $10^{-2}$& 1.80 x $10^{-2}$ \\
    5 & 1094.6$\pm$176.7 & 4.8 x $10^{-3}$ & 7.5 x $10^{-3}$ & 1.22 x $10^{-2}$& 2.84 x $10^{-2}$ \\
    \midrule
    \SetCell[c=6]{c}{\textbf{Cold Shroud Configuration}}\\
    \midrule
    6 & 1082.7$\pm$173.2 & 4.8 x $10^{-3}$ & 5.3 x $10^{-3}$ & 1.01 x $10^{-2}$& 3.46 x $10^{-1}$ \\
    7 & 1090.7$\pm$171.5 & 3.9 x $10^{-3}$ & 4.9 x $10^{-3}$ & 8.8 x $10^{-3}$& 3.20 x $10^{-2}$ \\
    8 & 1085.5$\pm$170.9 & 3.9 x $10^{-3}$ & 4.8 x $10^{-3}$ & 8.7 x $10^{-3}$& < CIC\\
    9 & 1093.1$\pm$170.8 & 4.4 x $10^{-3}$ & 6.1 x $10^{-3}$ & 1.06 x $10^{-2}$& < CIC\\
    10 & 1088.7$\pm$170.7& 4.8 x $10^{-3}$& 7.5 x $10^{-3}$& 1.22 x $10^{-2}$& < CIC\\
    \bottomrule 
\end{tblr}
\end{table}

\subsubsection{Parallel and Serial CIC}

 The parallel and serial CIC measured in the cold and warm shroud configuration are similar. The CIC is also similar for the measurements at different detector temperatures as the EMGain was kept fixed at $\sim$ 1000 \electron{}/\electron{} across these measurements. The average value of CIC across all dark images in both configurations is
\( 1.01 \times 10^{-2} \pm 1.37 \times 10^{-3} \, \text{e}^-/\text{pix/frame} \). While the average values of the parallel and serial CIC are \( 5.72 \times 10^{-3} \pm 1.04 \times 10^{-3} \, \text{e}^-/\text{pix/frame} \) and \( 4.41 \times 10^{-3} \pm 5.63 \times 10^{-4}\,\text{e}^-/\text{pix/frame} \), respectively. The CIC depends on the clocking speed, clock amplitude, the detector operation mode (inverted or non-inverted), and clock synchronization \cite{Tulloch_2011,Bush_2021}. The 1 MHz HV clock used for this work produces higher CIC compared to a 10 MHz clock \cite{Kyne_2020} because the slower clock allows for more spurious charges to be generated \cite{Daigle_2009,Daigle_2010}.

\subsubsection{Dark current in the two configurations}
\begin{figure}[ht]
    \centering
    \includegraphics[width=0.743\textwidth]{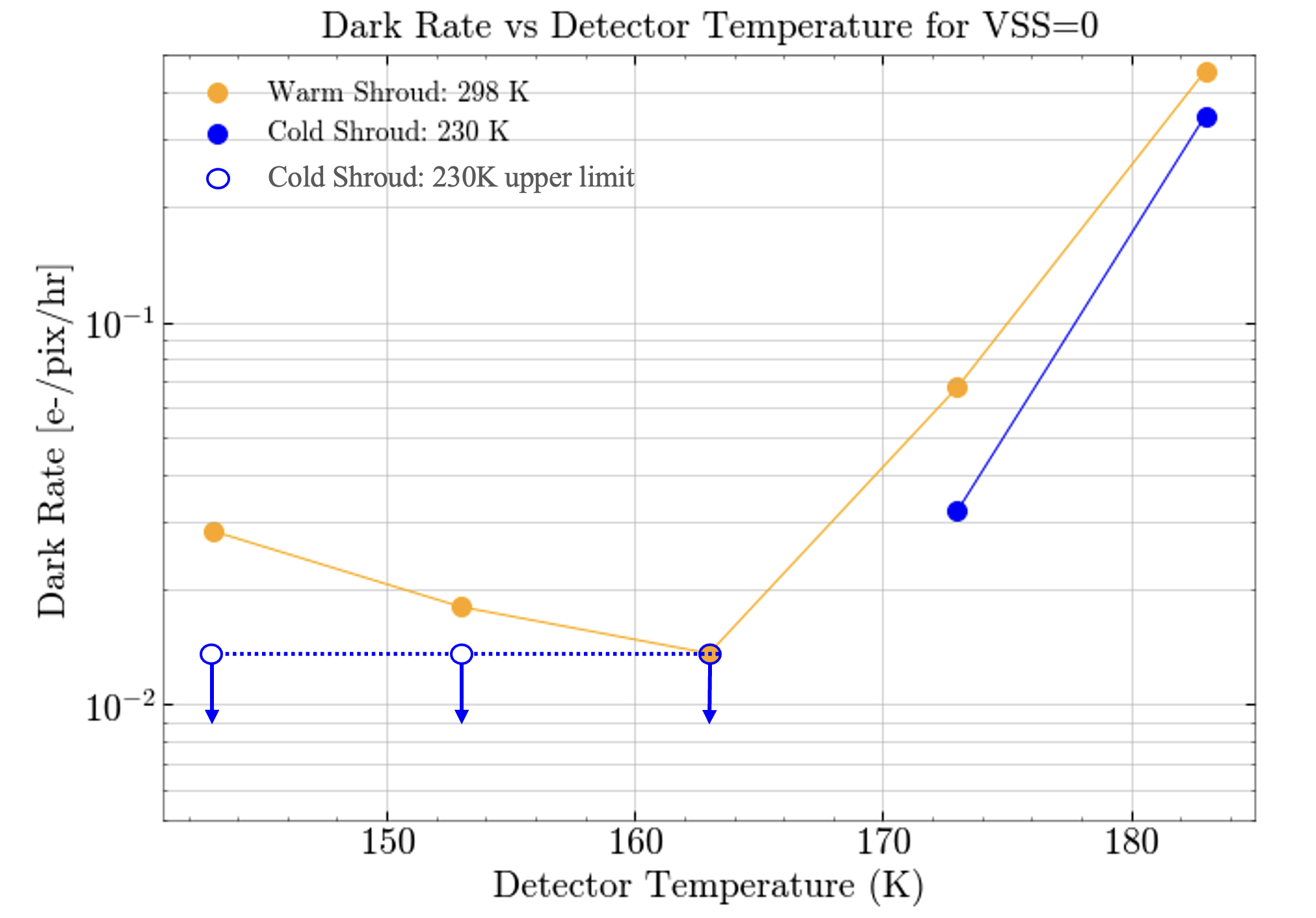}
    \caption{Preliminary Dark rate vs temperature measurements. The dark rate calculated at different temperatures is reported for measurements in the cold (Solid blue circles) and warm shroud configurations (Solid yellow circles). The lower limits dark rate are shown in the cold configuration where the dark rate could not be measured from the dataset taken for this work (Open blue circles). The upper limit is set based on the lowest measured dark rate in our data set- at 163K in the warm shroud configuration.}
    \label{fig:dark_temperature}
\end{figure}

The dark rate vs temperature plot for the two configurations is shown in Figure \ref{fig:dark_temperature}. In the warm shroud configuration, the dark current drops to a plateau below 163 K. This trend is similar to the dark plateau reported by Kyne et. al.\cite{Kyne_2020} 
In the cold shroud configuration, we found that at detector temperatures of 183 K and 173 K the dark rates are lower than the warm shroud configuration. However, for detector temperatures at 163 K and below, it was not possible to estimate the dark rate as the slope of the dark current vs exposure time was not detectable above the CIC. An example of this non-detection of dark rate is shown in Figure \ref{fig:dark_ptc_1}. At 183 K and 173 K, the dark rate in the cold shroud configuration is a factor of $\sim$ 30 \% and $\sim$  210 \%, respectively, lower than the corresponding measurements in the warm dark configuration. This reduction is potentially due to the reduction in the photons from the ambient environment reaching the detector in the cold shroud configuration. However, further investigation is needed to quantify this contribution in the plateau region. 

 A hurdle in making such a quantitative estimate is to understand the contribution of smearing and ensure that the smear corrections at lower detector temperatures are reliable. A visual inspection of images at 143 K and 153 K indicates that smear corrections are going to be challenging at these detector temperatures as the smearing appears to produce overlap between nearby events in the same row. As the CIC estimates at temperatures are similar it is possible that we could get better estimates of the dark current rate at lower detector temperatures if the smearing is properly accounted for. Further measurements (see Sec. \ref{sec:current_future}) and analysis are required to explain the non-detection of the dark rate in the cold shroud configuration for the low detector temperatures.

\section{Current Status and Future work}
\label{sec:current_future}

We are currently taking data for different VSS settings ranging from 0-6 V with the 1 MHz board. Significant efforts were made to further optimize the readout sequence based on the initial measurements presented here. We are working on improving our data reduction process to develop desmearing strategies for images taken at low detector temperatures. The ongoing measurements include additional dark images with longer exposure durations (up to 1200 seconds). This would add more data points to the Dark-PTCs and allow measurement of the dark rate at lower temperatures in the cold shroud configuration. The longer exposures have more pixels affected by cosmic ray events and tails. Future work includes testing different delta-doped and engineering-grade detectors with boards configured for both 1 MHz and 10 MHz HV clocks.

\acknowledgments 
This research is supported partly by the NASA Jet Propulsion Laboratory Strategic University Research (SURP) Grant between NASA JPL and the University of Arizona (JPL SURP FY21 SP21011). The research was carried out in part at the Jet Propulsion Laboratory, California Institute of Technology, under a contract with the National Aeronautics and Space Administration (80NM0018D0004). The authors would like to acknowledge the support from the NASA Nancy Grace Roman Technology Fellowship (80NSSC20K0684). Aafaque Khan is a NASA future investigator (FI) on the NASA Future Investigators in NASA Earth and Space Science and Technology (FINESST) grant (80NSSC21K2050). A.R.K. would like to acknowledge Dr. George Rieke, Dr. Ewan Douglas, Manny Montoya, Mitch Nash, Mario Rascon, Barry Davenport, Paul Arbo, John Ford, Dave Beaty, Frank Gacon, Elijah Garcia, Daniel Truong, Simran Agarwal, and Jessica Li at Steward Observatory for their valuable support and inputs. 
 
\bibliography{report} 
\bibliographystyle{spiebib} 

\end{document}